\documentclass[aps,prd,nofootinbib,showpacs,superscriptaddress,preprint]{revtex4-1}
\pdfoutput=1
\usepackage[utf8]{inputenc}
\usepackage{amsmath,amssymb}
\usepackage{graphicx}
\usepackage{multirow}
\usepackage[usenames,dvipsnames]{color}
\usepackage{float}
\usepackage[colorlinks,citecolor=blue]{hyperref}
\usepackage[compat=1.0.0]{tikz-feynman}
\usepackage{color}
\usepackage{subfigure}
\DeclareUnicodeCharacter{2212}{-}
\newcommand{\orcid}[1]{\href{https://orcid.org/#1}{\resizebox{10px}{!}{\includegraphics{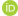}}}}

\begin{document}
%\preprint{IP/BBSR/2015-4}
\title{The effect of loop quantum gravitational rainbow functions on  the formation of naked singularities}

\author{Moh Vaseem Akram\orcid{0000-0001-7606-6717}}
\email{moh158947@st.jmi.ac.in}
\affiliation{Department of Physics, Jamia Millia Islamia, New Delhi, India}

\author{Imtiyaz Ahmad Bhat\orcid{0000-0002-2695-9709}}
\email{imtiyaz@ctp-jamia.res.in}
\affiliation{Center for Theoretical Physics, Jamia Millia Islamia, New Delhi, India}
\affiliation{Department of Physics, Central University of Kashmir, Ganderbal 191201 India}
\author{Anver Aziz\orcid{0000-0003-1762-7558}} 
\email{aaziz@jmi.ac.in}
\affiliation{Department of Physics, Jamia Millia Islamia, New Delhi, India}

\author{Mir Faizal\orcid{0000-0002-3292-9426 }}
\email{mirfaizalmir@gmail.com}
\affiliation{Canadian Quantum Research Center 204-3002, 32 Ave Vernon, BC V1T 2L7 Canada}
\affiliation{Irving K. Barber School of Arts and Sciences, University of British Columbia - Okanagan, Kelowna, British Columbia V1V 1V7, Canada and}
\affiliation{Department of Physics and Astronomy, University of Lethbridge, Lethbridge, AB T1K 3M4, Canada.}

\begin{abstract}
    In this paper, we will investigate     the consequences of  loop quantum gravitational modifications  on the formation of  naked singularities. The  loop quantum gravitational effects  will be incorporated in the collapsing system using gravity's rainbow. This will be done by using   rainbow functions, which are constructed from  loop quantum gravitational modifications to the    energy momentum dispersion. 
    It will be observed that this  modification will prevent the formation of  naked singularity.  Thus, such a modification can ensures that the   weak cosmic censorship hypothesis will hold for any collapsing system.
\end{abstract}

\maketitle
\section{Introduction}
In the  loop quantum gravity field variable is a self-dual connection, instead of the metric, and this new field variable is called the Ashtekar connection ~\cite{a1, a2}. The    Wheeler-DeWitt equation is then    reformulated in terms of   the traces of the holonomies of the Ashtekar connection ~\cite{b1, b2}. In loop quantum gravity, the area and volume  are represented by operators with discrete eigenvalues ~\cite{c1}. This discretization occurs near Planck scale, and this leads to a modification of the usual energy momentum dispersion relation   to a   modified dispersion relation  at Planck scale ~\cite{lqg1, lqg2}. Even though this  modified dispersion relation reduces to the usual dispersion relation in the IR limit, it considerably  deviates  from the usual energy momentum dispersion relation in the UV limit. This behavior of modification in the UV limit is also observed in the Horava-Lifshitz gravity, which  is motivated by Lifshitz scaling  between space and time ~\cite{12a, 12b}. In fact, such  Lifshitz scaling also  produces  a  deformation  of the standard energy momentum dispersion relation to modified dispersion relation in the UV limit of the theory ~\cite{12d, 12e}. The Lifshitz  deformation of    supergravity theories in the UV limit has also been studied ~\cite{lf1, lf2, lf4, lf5}.  It is also possible to motivate  a different form of    modified dispersion relation ~\cite{6, 7, 6ab, 7ab} using    the high energy cosmic ray anomalies  ~\cite{1, 2}.   The modification of standard dispersion relation to     modified dispersion relation  has motivated the construction of double special relativity, where the   Planck energy acts as another  universal constant ~\cite{4,5}. This theory of  double special relativity has been constructed using a    non-linear modification of the Lorentz group. Such modification to the 
dispersion relation  have also been obtained from  string field theory  ~\cite{7a,kuch,kuch1}.  
Thus, along with the phenomenological reasons, there are strong theoretical reasons to modify the energy momentum dispersion relation ~\cite{6, 7, 6ab, 7ab}.

It is possible to generalize double special relativity to curved space-time, and the resultant theory is called   gravity's rainbow ~\cite{gr, gr12}. In this theory  the metric depends on the energy of the probe used to analyze the geometry.  The energy dependence is introduced into the metric using rainbow functions.   As these rainbow functions depend on the energy of the probe,  which in turn depend implicitly on the coordinates, they cannot be  removed   by rescaling ~\cite{gr14, gr14ab, gr17ab, gr18ab}. In fact, this is expected as the gravity's rainbow is  related to Lifshitz deformation of geometries ~\cite{gr14}. Here the energy of the probe is converted into the length scale at which the probe is investigating the geometry, and this in turn has upper bound. Hence in gravity's rainbow, we cannot probe the space-time below Planck scale, as it is not possible to obtain an energy greater than the Planck energy. The experimental constraints on the rainbow functions from various experiments have been proposed ~\cite{w1}. The effect of these rainbow  functions on the black information paradox have also been investigated ~\cite{w2}. Such a modification of a higher curvature gravity  from gravity's rainbow has been studied, and used to analyze its quasinormal modes ~\cite{w5}. The gravity's  rainbow geometries have been used to study the modification to the physics of neutron stars ~\cite{w7}. In all these systems, the gravity's rainbow only changes the  UV Planck scale behavior of the system. However, the system behaves as the original un-deformed system, in the IR limit. Furthermore, the effect of rainbow function on a system depends on the kind of rainbow functions used to deform it ~\cite{w1}. Thus, in this paper, we will use  the   rainbow function  obtained from the   modification of energy momentum dispersion relation due to  loop quantum gravity ~\cite{y1, y2}.

It has been proposed that naked singularity will not form due to loop quantum gravitational effects, and this was done by analyzing the  non-perturbative semi-classical modifications to a collapsing system near the singularity ~\cite{singu}. Here we will demonstrate that these results can be obtained by analyzing the modification of energy momentum dispersion relation from loop quantum gravity ~\cite{gr, gr12}. To  analyze such effect from loop quantum gravitational modifications on the formation of naked singularity, we will use gravity's rainbow. It has been  proposed that naked singularities cannot form due to the weak   cosmic censorship conjecture ~\cite{cc12, cc14}. However, several violation of cosmic censorship conjecture  have been studied, and thus it seems that it is possible to form naked singularity in space ~\cite{cc16, cc18, cc19, cc20}. In fact, it has been argued that accretion properties of a collapsing system can  distinguish between a naked singularity, a wormhole and a black hole ~\cite{si12}. It has been suggested that  a naked singularity can form during  a  critical collapse of a scalar field ~\cite{si14}.   The  gravitational lensing by a strongly naked null singularity has been investigated  ~\cite{si15}. It has been demonstrated that   the nature of this divergence are not logarithmic. It has also been suggested that the formation of a naked singularity can be tested using astrophysical   observation  ~\cite{si17}. The shadow of a naked singularity without photon sphere has   been analyzed  ~\cite{si16}.   It has been argued  that the naked singularity cannot form due to quantum gravitational effects ~\cite{si18, si19}. Thus, it becomes important to analyze the effect of loop quantum gravity on the formation of naked singularities. As such modification to black hole geometries have already been studied ~\cite{gr14, gr14ab, gr17ab, gr18ab}, we will use gravity's rainbow to analyze the effect of loop quantum gravity on the formation of naked singularities. The effect of rainbow functions on the formation of naked singularities has also been studied, and it was observed that rainbow deformation can violate the cosmic censorship conjecture ~\cite{cca12}. However, we will argue that this cannot be the case, as due to the  rainbow deformation, it is not possible to probe space-time below Planck scale. This presents the formation of naked singularities.  
 
\section{Collapse in gravity´s rainbow  }
To properly analyze the formation of naked singularities, and weak cosmic censorship conjecture, we will first analyze the deformation of a solution with Einstein equation by rainbow functions, which are consistent with loop quantum gravity ~\cite{lqg1, lqg2}. It has been proposed that the  Einstein equations depend on the energy due to rainbow deformation,  $ {G_{\mu \nu } (E / E_p)=\mathcal{R}_{\mu\nu}(E / E_p)-\frac{1}{2}g_{\mu\nu}(E / E _p)\mathcal{R}(E / E_p)=8{\pi T}_{\mu \nu }}$ (with the standard unit $G=c=1$). Here   $E_p$ is Planck energy, and $E$ is the energy at which the system is probed. Thus, the geometry depends on the energy used to probe it. However, for $E<< E_p$, we can neglect the rainbow deformation.   It is possible to incorporate this  rainbow deformation into  the spherically symmetric line element  in   comoving coordinates $(t, \; r,\; \theta, \; \phi)$ as (with (-,+,+,+) as  the signature)
\begin{align}\label{1}
    ds^{2}= -\frac{e^{2 \lambda (t,r)}}{f(E)^2} dt^2 +\frac{e^{2 \psi (t,r)}}{g(E)^2} dr^2 +\frac{R(t,r)^{2}}{g(E)^2} d\theta^{2} +\frac{R(t,r)^{2} \sin^{2}{\theta}}{g(E)^2} d\phi^{2}
\end{align}
 where  $R(t,r)$ is the physical radius at time $t$ of the shell labeled by $r$. Here  $f(E)$ and $g(E)$ are rainbow  functions which make the metric depend on the energy of the probe $E$.  As in the IR limit, the  gravity's rainbow has to reduce to the usual general relativity, we expect that these rainbow function satisfy  
\begin{align}
    \lim_{E / E _p\to 0} f(E/ E _p) &= 1 &  \lim_{E /E
_p \to 0} g(E / E_p) & = 1
\end{align}

It is possible to obtain a specific form of these  rainbow function from loop quantum gravity  ~\cite{lqg1,lqg2}. We will use these these specific rainbow functions, and demonstrate that they will prevent the formation of naked singularities. However, before that we observe that any rainbow functions will limit the scale to which we can probe the system. It is possible to translate the uncertainty $\Delta p \geq 1/ \Delta x$ into a bound on energy of the probe $E$, as $E \geq 1/\Delta x$. Here $\Delta x$ would correspond to the scale to which any length scale in the system can be measured. We cannot take this length to be the same order as the Planck length, as such a bound is restricted by black hole physics ~\cite{uncer1, uncer2,quant}. 
As there is this minimal length in the system, we have a bound on the maximum energy need to probe such a minimal length. Now if  $E< E_p$ is such a maximum energy needed to probe, then we have to analyze rainbow modification to the systems when   $E$ is of the same order as $E_p$ (and can be neglected for $E<< E_p$). 
% As the    physical aerial radius is used to probe the system, we can observe that the bound can be expressed using $\Delta x \sim R(t,r)$, as the uncertainty can at most be of the same order as the physical aerial radius. Thus, we observe that $E \sim 1/ R(t,r) $. 
% Now it is not possible for to have Planck energy in gravity's rainbow $E < E_p$ \cite{gr, gr12}, and thus, we observe that $1/ C $ is bounded by a lowest possible value. In fact, we can explicitly obtain this lower possible value. 
 %We  observe that if the probe has Planck energy $E_p$, then it will probe Planck scale $L_p$. Thus, we can write $R(t,r) > L_p$, due to rainbow deformation. This bound on $R(t,r)$ imposed due to a deformation by rainbow gravity will prevent the formation  of naked singularities, when the rainbow functions are motivated by loop quantum gravity  
 %\cite{lqg1, lqg2}.   
 
The formation of naked singularity from a collapsing spherically symmetric object has been already been studied ~\cite{Singh:1994tb, Singh:1997wa}. Here we will investigate the rainbow deformation of a collapsing spherically symmetric object. 
 To explicitly analyze such a system, we express the energy momentum tensor for a spherically symmetric object in comoving coordinates as $T^{\mu}_{~{\nu}}=\text{diag}(-\rho,\; p_r, \; p_\theta, \;p_\theta),
$ where $\rho,\; p_r\; \text{and} \; p_\theta,$ are functions of $t$ and $r$.
After solving for different non-zero components of Einstein equations, and suitably deforming it by rainbow functions (using the $OGRE$ package ~\cite{Shoshany:2021iuc}), we get the following set of equations, 
\begin{align}
    G^t{}_t &= -\frac{g(E )^2\left(1-e^{-2\psi }\left(R'^2+2R R''-2R R'\psi            '\right)\right)+e^{-2\lambda }f(E )^2\left(\dot{R}^2+2
    R \dot{R}\dot{\psi }\right)}{R^2} \nonumber \\ & = -8\pi \rho \label{4.1} \\
    G^r{}_r &= -\frac{e^{-2\psi }g(E )^2}{R^2}\left(e^{2\psi }-R'^2-2R R'\lambda '\right)-\frac{e^{-2\lambda }f(E )^2}{R^2}\left(\dot{R}^2+2R
    \ddot{R}-2R \dot{R}\dot{\lambda }\right) \nonumber \\& =8\pi p_r \label{5.2} \\
    G^{\theta }{}_{\theta } &= G^{\phi }{}_{\phi }=\frac{e^{-2\psi }g(E )^2}{R}\left(R''+R'\left(\lambda '-\psi '\right)+R\left(\lambda ''+\lambda
    '^2-\lambda '\psi '\right)\right) \nonumber \\& -\frac{e^{-2\lambda }f(E )^2}{R}\left(\ddot{R}-\dot{R}\left(\dot{\lambda }-\dot{\psi }\right)+R\left(+\ddot{\psi
    }+\dot{\psi }^2-\dot{\lambda }\dot{\psi }\right)\right)\nonumber \\& = 8\pi p_\theta \label{8} \\
    G^r{}_t& = \frac{2e^{-2\psi }g(E )^2\left(\dot{R}\lambda '+R'\dot{\psi }-\dot{R}'\right)}{R}=0\label{7}
\end{align}
where $dot$ and $prime$ are the derivatives with respect to time and radial coordinates respectively. Now using  the definition of  Misner–Sharp Mass~$F(t,r)$ as $ 1-(F(t,r)/R)=g^{\mu \nu }\triangledown _{\mu } R \triangledown _{\nu }R$  ~\cite{Bambi:2019xzp}, we can write  a rainbow deformation of 
Misner–Sharp Mass as 
\begin{align}\label{4}
  F(t,r)=R\left(1+f(E )^2e^{-2\lambda }\dot{R}^2-g(E )^2e^{-2\psi }R^{\prime ^2}\right) 
\end{align}
It  would be useful to define $j=1-g(E)^2$. Using this, we observe $F'=j R'+8\pi  \rho  R^2R'$ and $\dot{F}=j \dot{R}-8\pi  p_rR^2\dot{R}$. Now due to  the  conservation of energy momentum tensor ${T^{\mu }{}_{\nu ;\mu }=0}$, we obtain $\dot{\rho }+\left(\rho +p_r\right)\dot{\psi }+{2\left(\rho +p_{\theta }\right)\dot{R}}/{R}=0$ and $p_{\theta }'+\lambda '\left(\rho +p_r\right)+{2\left(p_r-p_{\theta }\right)R'}/{R}=0$. 
The  acceleration equation can be obtained  by defining $h(t,r)=1-g(E)^2e^{-2\psi }R^{\prime ^2}$. Here   $h(t,r)$  depends on the rainbow function $g(E)$, hence is an energy dependent  function. Thus, the   initial density and velocity profile conditions of the collapsing dust ~\cite{Singh:1994tb} would depend on the maximum energy of the system, and would get deformed in the UV limit. We observe that using $h(t,r)$, we can obtain,  
\begin{align}\label{12}
    {F(t,r)}/{R}=f(E )^2e^{-2\lambda}\dot{R}^2+h(t,r)
\end{align} 
and ${\dot{h}}/{(1-h)}=- {2\dot{R}\lambda '}/{R'}$. Now we can write    the equation for acceleration as 
\begin{align}\label{14}
   \ddot{R}&=\dot{R}\dot{\lambda }+\frac{e^{2\lambda }}{f(E )^2}\left(\frac{j }{2R}-4\pi  R p_r-\frac{F}{2R^2}+\frac{(1-h)}{f(E
)^2R'}\left(-\frac{p_r'}{\rho +p_r}+\frac{2R'\left(p_{\theta }-p_r\right)}{R\left(\rho +p_r\right)}\right)\right)
\end{align}
We can  define the proper time as ${{d\tau } =\frac{e^{\lambda }}{f(E)}}{dt} $, and use it to obtain 
\begin{align}\label{16}
    \frac{d^2R}{d \tau ^2} &= \frac{\left(1-f(E )^2\right)\dot{R}\dot{\lambda }}{e^{2\lambda }}+\frac{1}{f(E )^2}\bigg(\frac{j}{2R}-4\pi  R p_r-\frac{F}{2R^2}+\frac{(1-h)}{f(E )^2R'}\bigg(-\frac{p_r'}{\rho +p_r}\nonumber\\&+\frac{2R'\left(p_{\theta }-p_r\right)}{R\left(\rho
+p_r\right)}\bigg)\bigg)
\end{align}
 This modified equation  depends on the maximum  energy of the system due to the rainbow functions $f(E)$ and $g(E)$. So,  the conditions for the collapse from this equation will be implicitly energy dependent. This interesting  energy dependence constraints can have important consequences for the formation    of  naked singularity.   

Let us take an example of  perfect fluid where the  radial and tangential pressures are equal  ($p_r=p_\theta=p$), and write the equation of acceleration for this perfect fluid as 
\begin{align}\label{17}
   \frac{d^2R}{d \tau ^2}=\frac{\left(1-f(E )^2\right)\dot{R}\dot{\lambda }}{e^{2\lambda }}+\frac{1}{f(E )^2}\left(\frac{j}{2R}-4\pi  R p-\frac{F}{2R^2}-\frac{(1-h)}{f(E )^2R'}\frac{p'}{\rho +p}\right)
\end{align}
 Now by  setting the  acceleration and velocity equal to zero,  we obtain the Oppenheimer–Volkoff equation for hydrostatic equilibrium ~\cite{Oppenheimer:1939ne}
\begin{align}\label{18}
    {-R^2p'=\frac{f(E )^2(\rho +p)}{1-\frac{F}{R}}\left(\frac{F}{2}+4\pi  R^3 p-\frac{R(1 - g(E)^2) }{2}\right)}
\end{align}
where we have used ${h={F}/{r}}$ (after putting velocity equal to zero). The derivative of radial pressure appears in the acceleration equation. The fate of the collapse is determined by the radial and tangential pressure. It is noted from Eq.\eqref{16} that positive tangential pressure and negative tangential  pressure  opposes  and supports the collapse,  respectively. The rainbow functions that appear in Eq.\eqref{14} and \eqref{16} could reduce the value of acceleration, but could not change the overall sign of these equations, because these rainbow functions are always less or equal to one ($f(E) \leq 1 \; \text{and} \; g(E) \leq 1)$). 
Thus, they can only change the relative magnitude of these forces, and the effect these forces will have on the collapsing system.

\section{Conditions for collapse from acceleration equation} 
 In this section, we will discuss   different cases for the spherical collapse, and the effect of the  rainbow functions on them.   For the dust case,   we set the tangential and radial pressures equal to be zero ($p_r=p_\theta=0$). So, let us  assume that collapse starts from rest at $t=0$,  and  we set the initial conditions  as $ 
      R(t,r)\big|_{t=0}=r ~ \text{and}
    ~  F(t,r)\big|_{t=0}=F_c(r)$. 
 Using these initial conditions in  expression of $\dot{F}$,  Misner- Sharp mass becomes
 \begin{align}
     F(t,r)=F_c(r) + (1-g (E)^2)(R-r) \label{18.1}
 \end{align}
 Here $ p_{\theta}= p_{r} =0$ for the dust case and  the $\lambda$ is only the function of $t$.  Thus, we can redefine $t$ and set $\lambda=0$. Similarly, for the dust case, $h(t,r)$ will be function of $r$ only. With these re-definitions of the variables, we can write the rainbow deformed  metric for the dust case as
  \iffalse
  From Eq.(\ref{13}) it is evident that $h(r)$  is a function of $r$ only. The metric (\ref{1}) with the help of Eq.(\ref{11}) can be written as,
  \fi
\begin{align}\label{20}
    {{ds}^2=\frac{-{dt}^2}{f(E )^2}+\frac{R^{{\prime 2}}}{1-h(r)}{dr}^2+\frac{R^2}{g(E )^2}{d\Omega }^2}
\end{align}
where $d\Omega^2=d\theta^2+sin^2\theta ~ d\phi^2$. Here $h(r)$ is less than $1$, so that  the coefficient of $dr^2$ remain spacelike. 

From Eq.(\ref{12}), we find the equations, governing the behaviour of dust in the framework of rainbow functions that are consistent with quantum gravity.
\begin{align}\label{21}
    \frac{F_c(r)+ j(R-r)}{R}=f(E)^2\dot{R}^2+h(r)
\end{align}
Eq.\eqref{21} is discussed in details in the next section.

For collapse to start, the acceleration has to be negative or inward. We will use this fact to derive the condition for collapse to discuss different cases of  tangential pressure, radial pressure and perfect fluid.
\subsection{Tangential Pressure}\label{tangential}
Now if the  tangential pressure $p_{\theta}$ is non-zero, and the  radial pressure $p_{r}$ is zero,  a  simplification of  Eq.(\ref{16}) occurs. This simplified equation of acceleration can be written as 
\begin{align}\label{22}
   \frac{d^2R}{d \tau ^2}=\frac{\left(1-f(E )^2\right)\dot{R}\dot{\lambda }}{e^{2\lambda }}+\frac{1}{f(E )^2}\left(\frac{j}{2R}-\frac{F}{2R^2}+\frac{(1-h)}{f(E )^2R'}\frac{2R'p_{\theta }}{\rho  R}\right)
\end{align}

For collapse to begin, the acceleration has to be negative. Assuming the collapse to start from rest($\dot{R}=0$), and from   Eq.(\ref{12}) we obtain  ${h=\frac{F_c}{r}}$,  then the condition for collapse to begin at $t=0$  turns out to be,
\begin{align}
  \frac{F_c}{2r}>\frac{\frac{j }{2}+\frac{2p_{\theta }}{f(E )^2\rho }}{1+\frac{4p_{\theta }}{f(E )^2\rho }}=p_\theta \; \text{condition}
  \label{23}
\end{align}

If the condition of Eq.\eqref{23} is satisfied at $t=0$ and holds at all later times, the collapse will begin. In this case, the  singularity will form  at $r=0$,  if acceleration is negative, throughout the later time evolution of the system. We also observe from  Eq.\eqref{23}, that this system depends on the   energy because of the   rainbow functions. 

\begin{figure}[htp]
    \centering
    \includegraphics[height=6cm,width=8cm]{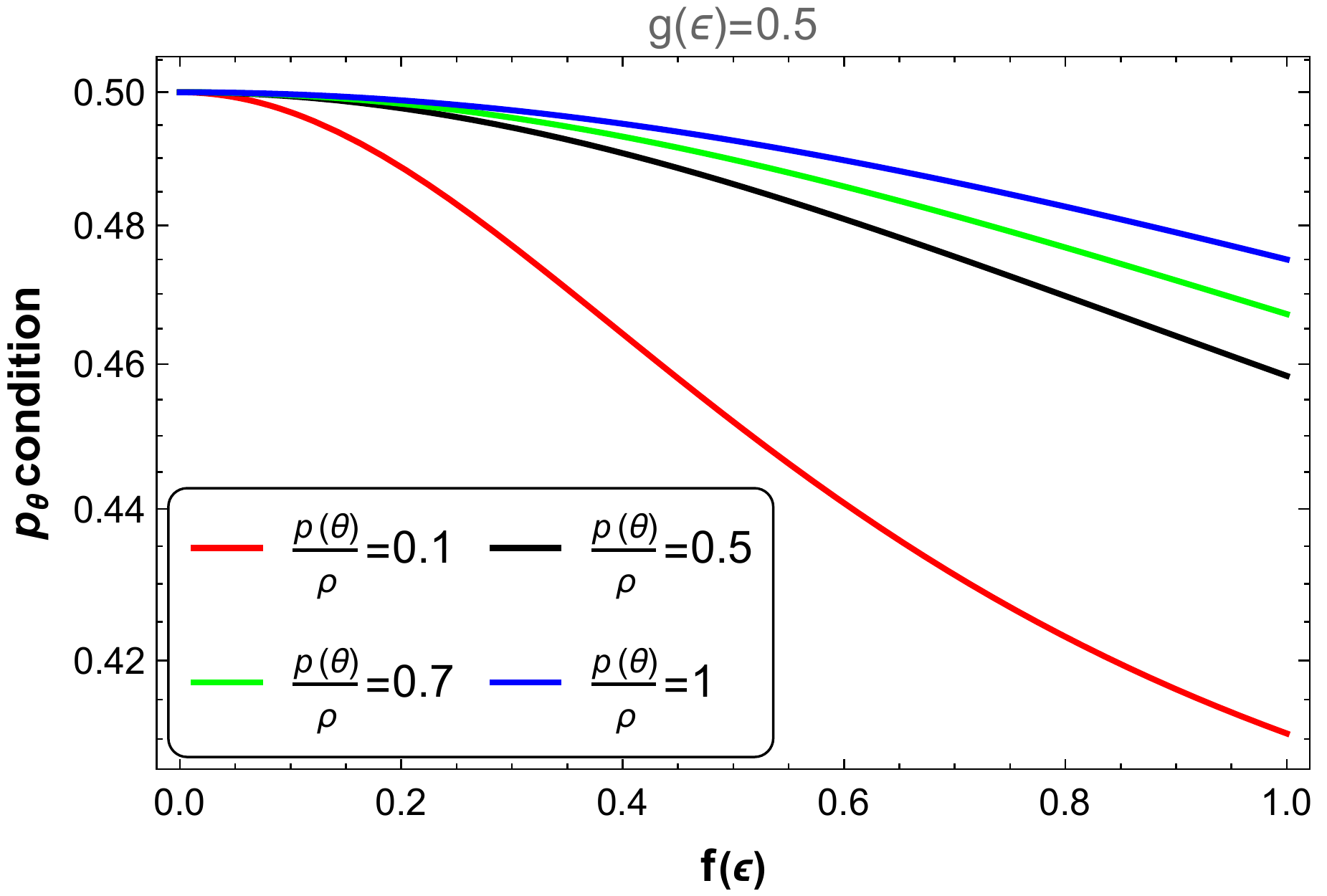}
    \includegraphics[height=6cm,width=8cm]{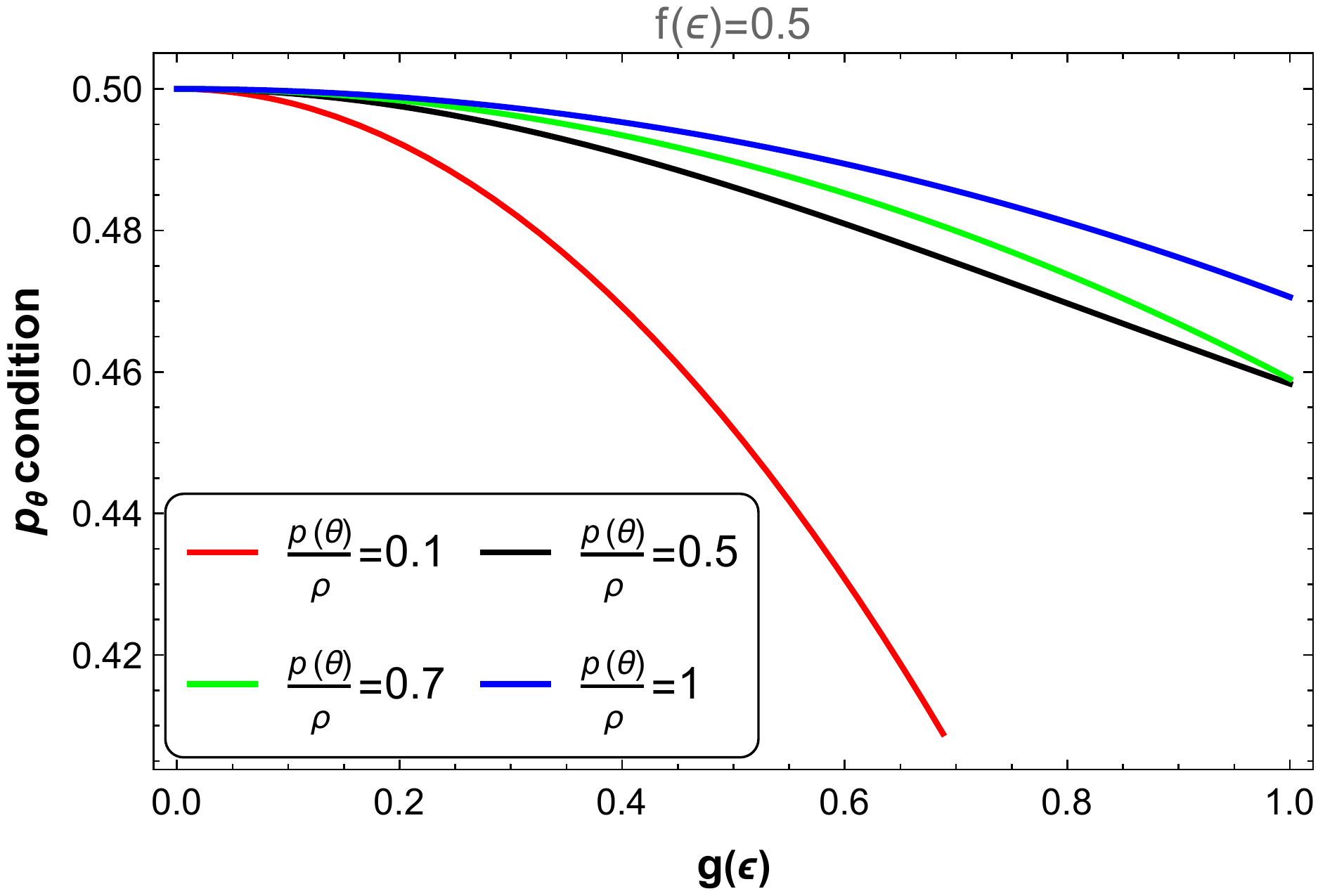}
    \caption{Plots of R.H.S. of Eq.\eqref{23} versus f($E)$  and g($E)$ for different of ratio of $p_\theta/\rho.$}
    \label{pfg}
\end{figure}
The dependence on the ratio $p_{\theta}/\rho$   generally evolves with time. However, to 
 examine the dependency of collapse on these rainbow functions, we assume the ratio  $p_{\theta}/\rho$, remains constant in  time. For positive tangential pressure and density, we assume that this ratio lies in the range  $0 \leq  p_\theta/\rho \leq 1$. In Fig.\eqref{pfg}, we have plotted Eq.\eqref{23}. It follows from both the graphs, as the energy of the system   tends towards the Planck scale,  $f(E)$ and $g(E)$ tends towards zero. This  makes it difficult for collapse to happen, which in turn   has consequences for the singularity formation. Hence we can conclude that the rainbow functions  modify the collapsing system.  In the IR limit,  $f(E) = g(E)=1$, we get back the same  condition as described in ~\cite{Barve:1999ph}. 
\begin{align}
    {\frac{F_c}{r}>\frac{4\left.p_{\theta }\right/\rho }{1+4\left.p_{\theta }\right/\rho }}
\end{align}

\subsection{Radial Pressure}
Here we can analyze the effect of a non-zero   radial pressure $p_{r}$, with a vanishing  tangential pressure $p_{\theta}$. This assumption again leads to a  simplification of  Eq.(\ref{16}), and this modified  equation of acceleration is  given by
\begin{align}\label{48}
    \frac{d^2R}{d \tau ^2} &= \frac{\left(1-f(E )^2\right)\dot{R}\dot{\lambda }}{e^{2\lambda }}+\frac{1}{f(E )^2}\bigg(\frac{j}{2R}-4\pi  R p_r-\frac{F}{2R^2}-\frac{(1-h)}{f(E )^2R'}\left(\frac{2R'p_r}{R\left(\rho +p_r\right)}+\frac{p_r'}{\rho +p_r}\right)\bigg)
\end{align}
Now  the condition   on radial pressure for the collapse to begin at $t=0$ turns out to be,
\begin{align}\label{49}
    \frac{F_c}{2 r}>\frac{\frac{j }{2}-4\pi  r^2 p_r-\frac{2p_r+r p_r'}{f(E)^2(\rho +p_r)}}{1-\frac{4p_r+2 r p_r'}{f(E)^2(\rho +p_r)}}
\end{align}
Again, in the IR limit $f(E) = g(E)=1$,    we obtain    the  condition for collapse which can be derived from general relativity ~\cite{Barve:1999ph}. This depends on  the density $\rho $, radial pressure and its derivative
\begin{align}
     \frac{F_c}{2 r}>\frac{4\pi  r^2 p_r+\frac{2p_r+r p_r'}{\rho +p_r}}{-1+\frac{4p_r+2 r p_r'}{\rho +p_r}}
\end{align}

\subsection{The Perfect Fluid}
In this approximation, called the   perfect fluid approximation both  the radial and tangential pressures are set to zero, $p_r = p_\theta = p$. In this case, Eq.(\ref{16}) becomes,
\begin{align}\label{24}
   {\frac{d^2R}{d \tau ^2}=\frac{\left(1-f(E )^2\right)\dot{R}\dot{\lambda }}{e^{2\lambda }}+\frac{1}{f(E )^2}\left(\frac{j}{2R}-4\pi  R p-\frac{F}{2R^2}-\frac{(1-h)}{f(E )^2R'}\frac{p'}{\rho +p}\right)}
\end{align}

For collapse to begin, the acceleration has to be negative. Now using   Eq.(\ref{12}), we obtain  ${h=\frac{F_c}{r}}$ and the condition for collapse to begin at $t=0$ turns out to be,
\begin{align}\label{24.01}
    {\frac{F_c}{2r}>\frac{\frac{j }{2}-4\pi  r^2 p-\frac{r p'}{f(E )^2(\rho +p)}}{1-\frac{2r p'}{f(E )^2(\rho +p)}}}
\end{align}
Here in the IR limit,  $f(E) = g(E)=1$, we obtain the condition which can be obtained from general relativity
\begin{align}
    \frac{F_c}{2r}>\frac{4\pi  r^2 p+\frac{r p'}{\rho +p}}{-1+\frac{2r p'}{\rho +p}}
\end{align}
 We have investigate  the dependence of this  conditions   for the different physical values of the pressure, such as when the  tangential pressure or  radial pressure are zero or  both are set equal.  We are interested in studying the case of dust and the effect of loop quantum gravitational modifications on the formation of naked singularity using gravity’s rainbow framework. 

\section{The Dust Solution}\label{dust}
The Tolman-Bondi dust collapse has been investigated in general relativity ~\cite{Singh:1994tb,Barve:1999ph,Singh:1997wa,Gundlach:1997wm}. The results for a marginally bound case and a non-marginally bound case have been analyzed in these studies.  However, we will restrict our discussion  here to only  the marginally bound case $i.e.~ h(r) = 0$. The results for the non-marginally bound case can be derived using the same procedure. Now  we will explicitly use the rainbow function motivated from loop quantum gravity ~\cite{y1, Amelino-Camelia:2008aez}
\begin{eqnarray}
     f(E )=1, &&\,\,\,\,\,\,\,\,\,\,\,\,\,\, g(E )=\sqrt{1-\eta  \frac{E}{E_p}}
\end{eqnarray}
Using these rainbow functions, we can explicitly write the metric as  
\begin{align}\label{29}
    {ds}^2=-{dt}^2+R^{{\prime 2}}(t,r){dr}^2+\frac{R^2(t,r)}{g(E )^2}{d\Omega }^2
\end{align}
Here $ {\dot{R}}$ will also depend on the energy of the probe 
\begin{align}\label{30}
    {\dot{R}=-\sqrt{j +\frac{F_c(r)-rj }{R}}}
\end{align}
We note that, along radial null geodesic, we can write 
\begin{align}\label{29.1}
    \frac{\partial t}{\partial r}=R'
\end{align}
Solving the above equation, at constant $r$ using the boundary condition $R_{t=0}=r$, we obtain 

\begin{align}\label{31}
   t&= \frac{\sqrt{r F_c}}{j }-\frac{F_c-rj}{j^{3/2}}\tanh ^{-1}\big(\sqrt{{F_c}}{(rj)^{-1}}\big)  -\frac{R}{j }\sqrt{j +\frac{F_c-rj}{R}}\nonumber\\&+\frac{F_c-rj}{j^{3/2}}\tanh
^{-1}\bigg(\sqrt{1+\frac{F_c-rj}{Rj}}\bigg)
\end{align}
Using the standard procedure, we introduce the auxiliary variables $u,X$  ~\cite{Dadhich:2003gw},
\begin{eqnarray}
    u=r^\alpha  ,\;  \alpha>0, &&
    \; X=\frac{R}{u}
\end{eqnarray}
In order for the singularity at $r = 0$ to be naked, radial null geodesics should be able to propagate outwards, starting from the singularity. A necessary and sufficient condition for this to happen is that the area radius $R$ increases along an outgoing geodesic, because $R$ becomes negative in the unphysical region. Thus in the limit of approach to singularity, we write 
\begin{align}\label{ghosh}
    X_0 &= \underset{R\to  0,u\to  0}{\lim}\frac{R}{u}=\underset{R\to  0,u\to  0}{{\lim}}\frac{{dR}}{{du}}
    \nonumber \\ & =\underset{R\to  0,r\to  0}{{\lim}}\frac{1}{\alpha r^{\alpha -1}}\frac{{dR}}{{dr}}=\underset{R\to  0,r\to  0}{{\lim}}\frac{1}{\alpha r^{\alpha -1}}\bigg(R'+\frac{\partial t}{\partial r}\dot{R}\bigg)\nonumber\\&=\underset{R\to  0,r\to  0}{{\lim}}\frac{1}{\alpha r^{\alpha -1}}R'\left(1+\dot{R}\right)
\end{align}
We can  evaluate $R'$ from Eq. \eqref{31} and  in the resulting expression, substitute $R=Xr^\alpha$. Then divide by $r^{\alpha-1}$, we obtain  the following expression
\begin{align}\label{32}
  \frac{R'}{r^{\alpha -1}}&=X\big\{-2r^{3/2} A_1 A_2F'_c\sqrt{j } +r \sqrt{F_c} \big(A_2+r^{\alpha /2} A_1\sqrt{X} \sqrt{j}  +2A_1A_2\big(\tanh ^{-1}\big(\sqrt{{F_c}{(rj)^{-1}  }}\big)\nonumber\\&-\tanh ^{-1}A_1\big) \big) \big(-j +F'_c\big)\big\}\big\{\sqrt{F_c} \big(r A_2j  +r^{\alpha /2} A_1\sqrt{X} \big(r-2 r^{\alpha } X\big)j^{3/2}
  \nonumber\\& +F_c \big(A_2-r^{\alpha /2} A_1\sqrt{X} \sqrt{j } \big)\big)\big\}^{-1}
\end{align}
where $A_1^2= {1-{r^{-\alpha } (j  r+F_c)}{(j  X)^{-1}}}   \; \text{and} \;
    A_2^2=j  (r-r^{\alpha } X )+F_c$. Assuming $u=r^\alpha$ along the radial null geodesic, we can write the following 
\begin{align}
    \frac{dR}{du}=\frac{1}{{\alpha r}^{\alpha -1}}\frac{{dR}}{{dr}}=\frac{1}{{\alpha r}^{\alpha -1}}\left(R'+\frac{\partial
t}{\partial r}\dot{R}\right)
\end{align}
Using the expression from Eq. \eqref{29.1}, the above equation takes the form,
\begin{align}\label{33}
    {\alpha \frac{{dR}}{{du}}=\left(1-\sqrt{j +\frac{F_c-rj }{R}}\right)\frac{R'}{r^{\alpha -1}}}
\end{align}
here $\frac{R'}{r^{\alpha -1}}$ is given by Eq. \eqref{32}. Roots analysis is done for the above case numerically as it is not possible analytically. To proceed further we consider the power series form of $F_c(r)$ as 
\begin{align}\label{seriesF}
    F_c(r)=F_0+F_1 r+F_2 r^2+F_3 r^3+ F_4 r^4...
\end{align}

\begin{figure}[H]
    \centering
    \includegraphics[scale=0.4]{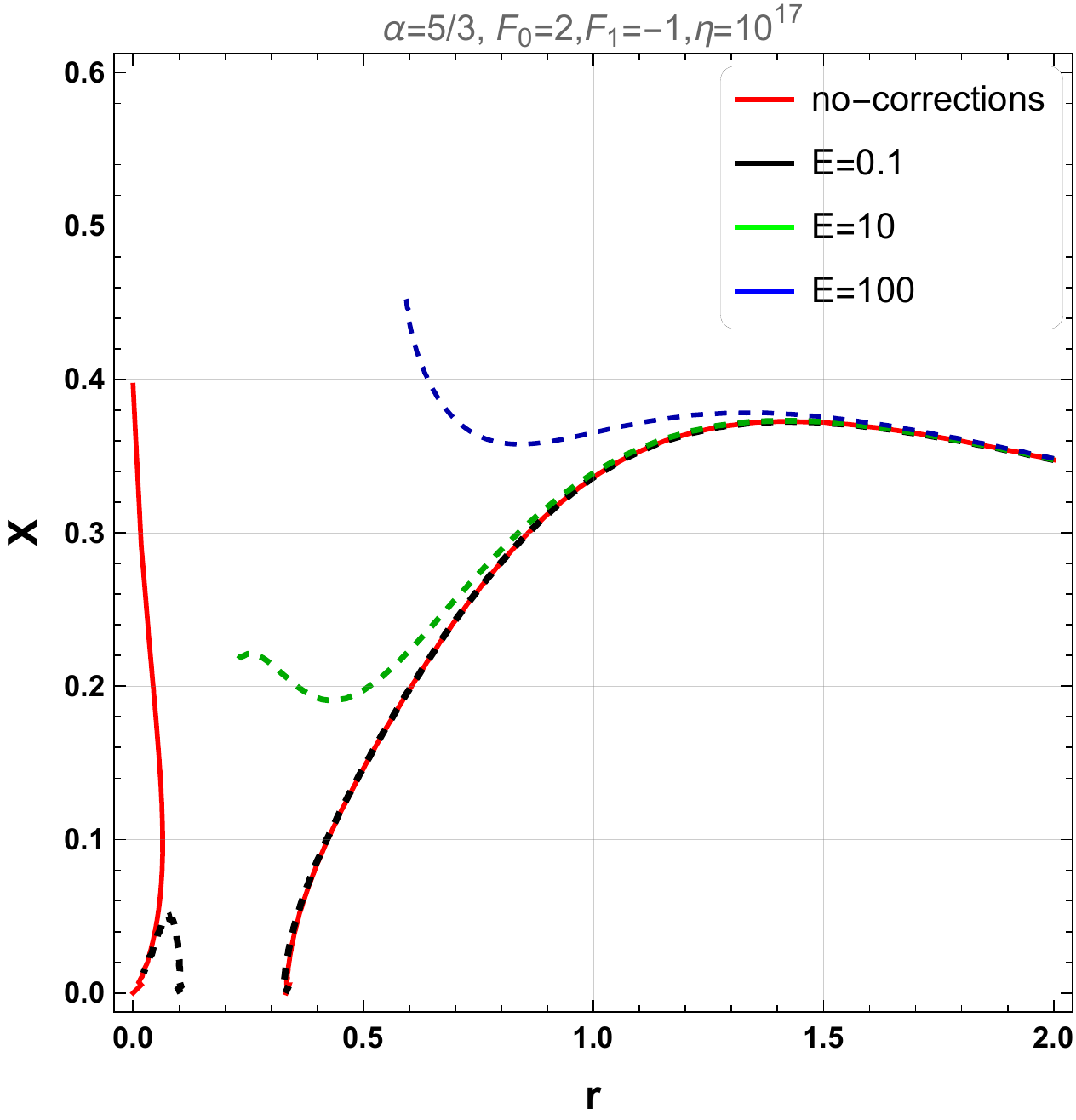}
    \includegraphics[scale=0.4]{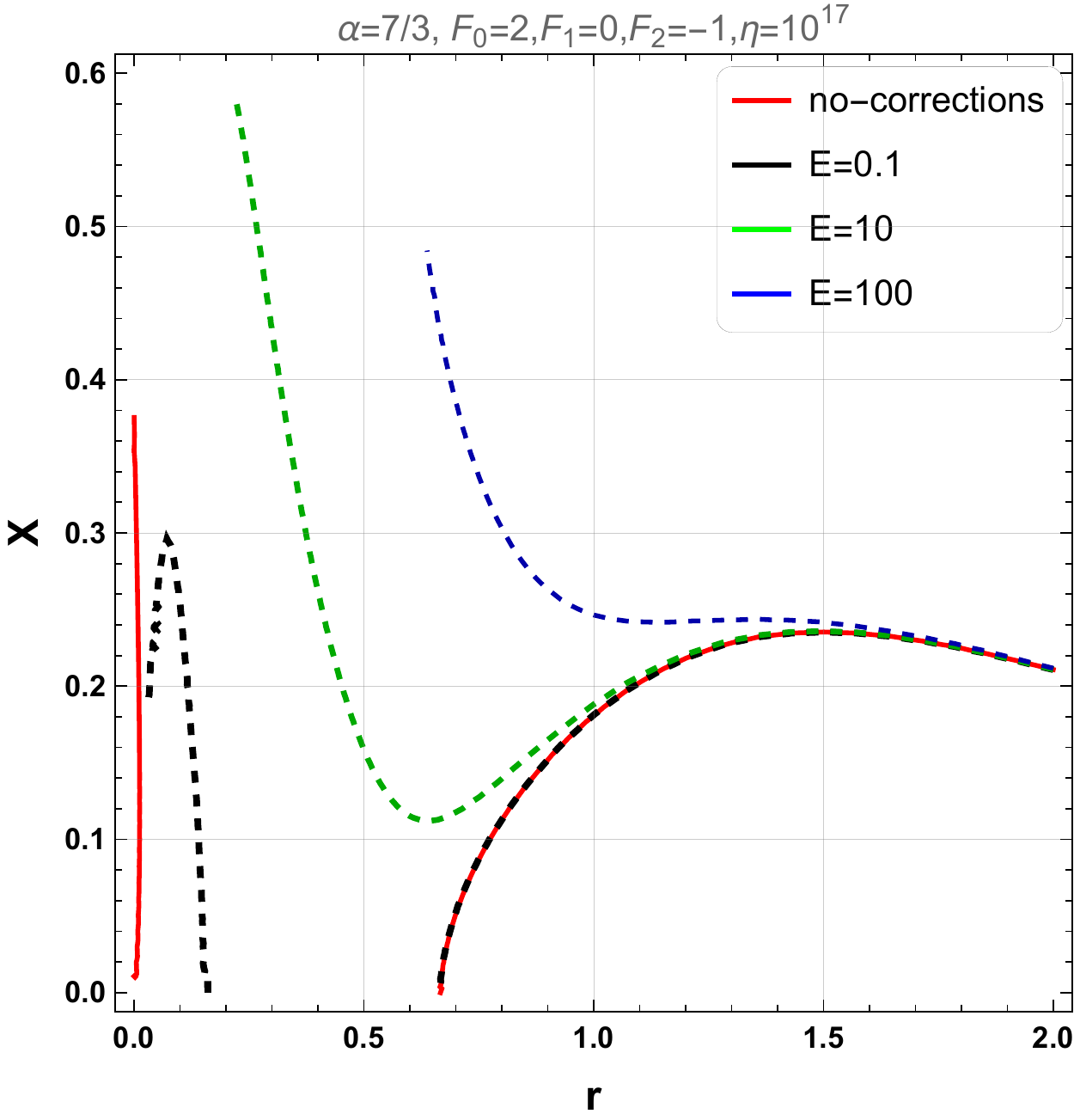}
    \includegraphics[scale=0.4]{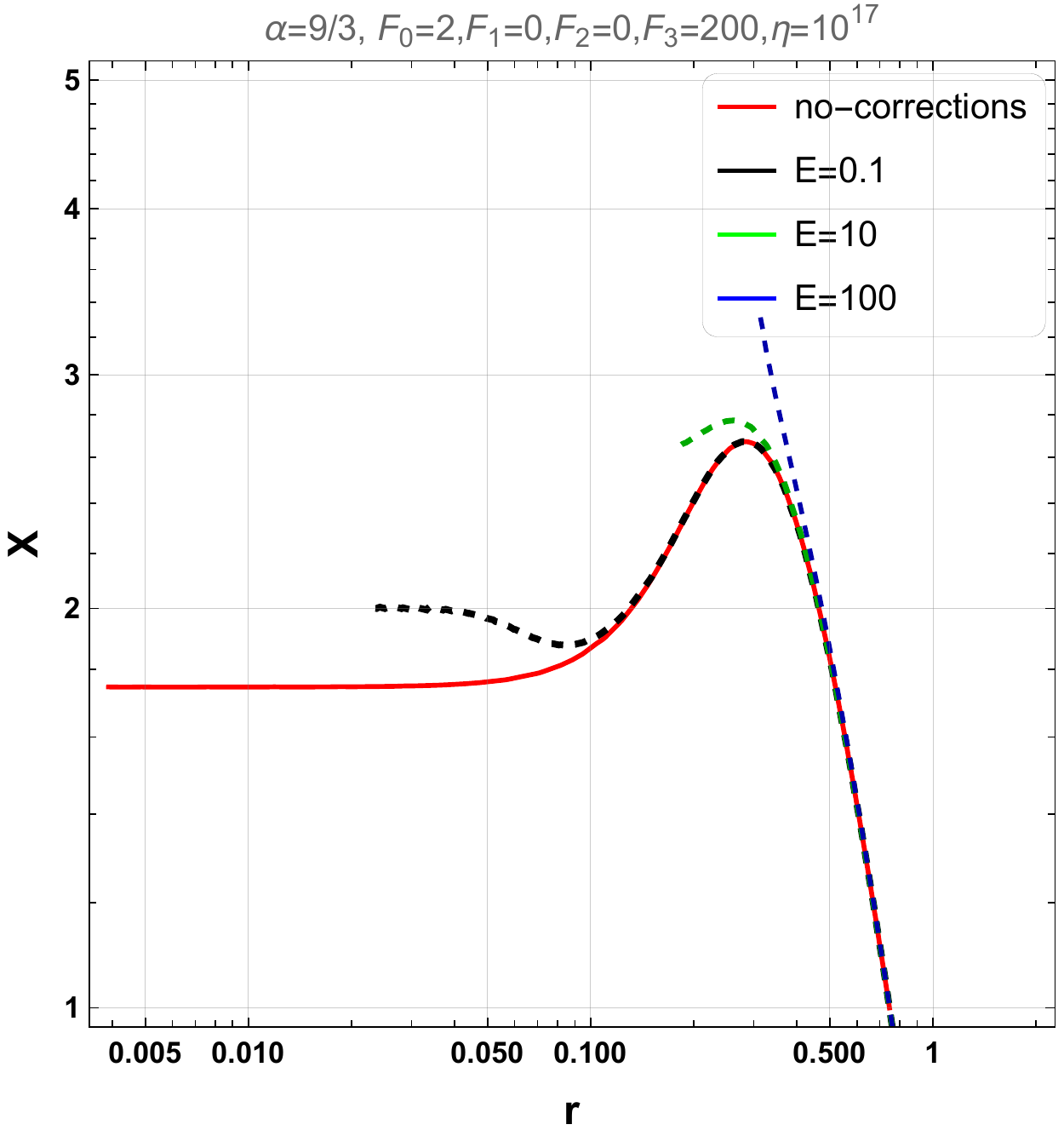}
    \caption{Plots of X versus r for various values of Energy $E=0,0.1,10$ and $100$ in Rainbow energy function $f(E )=1, g(E )=\sqrt{1-\eta  \frac{E}{E_p}}$. Here $E_p$ is taken as $10^{19}$ and $\eta=10^{17}$}
    \label{fig:Xr}
\end{figure}

We have plotted the value of $dR/du=R/u$ versus $r$ to see how the plot behaves near $r\to0$. This is done to investigate if $dR/du=R/u$ has a solution,  in the limit $r\to0$. As shown by red curves in  Fig. (\ref{fig:Xr}) that, in general relativity there is a real and positive value of $X$ in the limit $r\to 0$, while in gravity's rainbow, the curves departs farther from the vertical axis with increasing the value of $E$ as depicted from the plots itself. This suggest that  there is no real and positive value of $X$ exists as $r\to 0$ due to the deformation by gravity´s rainbow as opposed to general relativity. In fact, one can check  numerically that it gives the  complex roots of equation $dR/du=R/u$ for all values of $r<r_0$ near $r=0$. This can be checked for different values of the energy of the probe $E$. We observe as long as $E<E_p$, and of the same order (such that we cannot neglect $E/E_p$), the naked singularity will not form. This has been explicitly demonstrated for $E=0.1,10$ and $100$. For numerical analysis values of all model parameters are mentioned in plots . It can be concluded from above analysis that due to the deformation from  the rainbow function which are   motivated from loop quantum gravity ~\cite{y1, Amelino-Camelia:2008aez},    the    naked singularity will not form.

\section{Conclusion}

It is known that the energy momentum dispersion relation would be modified at Planck scale due to loop quantum gravitational effects. This  loop quantum gravitational modified energy momentum dispersion relation can be used to  obtain suitable rainbow  functions. We have  used those rainbow function to analyze the effect of loop quantum gravity on a collapsing system. We demonstrate that the  modifications to the collapsing system by loop quantum gravity prevents the formation of a   naked singularity. We comment that this was expected, as in loop quantum gravity the Planck scale structure of space-time is modified, and so we would expect the singularities would be removed due to these effects. This is explicitly demonstrated in this paper using  gravity's rainbow. The maximum energy of the system, which acts as a probe, is fixed using the uncertainty principle. Thus, the energy of the rainbow functions is expressed in terms of distance scale in the collapsing system. 

It would be interesting to analyze the collapse in different modified theories of gravity using this Planck scale modification. Thus, we can analyze this system  gravity with higher curvature terms, and then deform that system by gravity's rainbow. We would also like to point out that the deformation by gravity's rainbow depends on the rainbow functions. Here the rainbow functions were obtained using results from loop quantum gravity. However, it is possible to obtain rainbow functions from other motivations.  It is expected that the formation of naked singularities  would also depend critically on the kind of rainbow functions used to deform the system.  

\section{Acknowledgement}
M. V. Akram would like to thank Sukratu Barve for useful discussions.  I.A.Bhat would like to thank the Centre for Theoretical Physics, Jamia Millia Islamia, New Delhi for its hospitality where a major part of this work was carried out.

\end{document}